\documentclass[a4paper,fleqn]{cas-sc}

\usepackage[authoryear,longnamesfirst]{natbib}
\usepackage{xcolor} 
\usepackage{subcaption}
\usepackage{soul}

\usepackage{lineno}

\def\tsc#1{\csdef{#1}{\textsc{\lowercase{#1}}\xspace}}
\tsc{WGM}
\tsc{QE}

\usepackage{wasysym}

\usepackage{ifthen}

\newboolean{showcomments}
\setboolean{showcomments}{true} 

\newcounter{comments}

\newcommand{\commentby}[3]{%
  \ifthenelse{\boolean{showcomments}}{%
    \addtocounter{comments}{1}%
    {\color{#1}\bfseries[#2 comment \thecomments: #3]}%
  }{}%
}

\begin{document}
\let\WriteBookmarks\relax
\def\floatpagepagefraction{1}
\def\textpagefraction{.001}

\shorttitle{Escape of satellite ejecta}    

\shortauthors{Castro-Cisneros, Malhotra, Rosengren}  

\title [mode = title]{Analytical estimates for heliocentric escape of satellite ejecta}  



%

\author[1]{Jose Daniel Castro-Cisneros}[
    orcid=0000-0002-6624-4214
]

\cormark[1]


\ead{jdcastrocisneros@arizona.edu}


\credit{Conceptualization, Formal analysis, Investigation, Visualization, Writing - original draft, Writing - review \& editing}

\affiliation[1]{organization={Physics Department, The University of Arizona},
            addressline={1118 E. Fourth Street}, 
            city={Tucson},
            postcode={85721}, 
            state={AZ},
            country={USA}}

\author[2]{Renu Malhotra}




\credit{Conceptualization, Funding acquisition, Formal Analysis, Supervision, Writing - original draft, Writing - review \& editing}

\affiliation[inst2]{organization={Lunar and Planetary Laboratory, The University of Arizona},
            addressline={1629 E. University Blvd.}, 
            city={Tucson},
            postcode={85721}, 
            state={AZ},
            country={USA}}


\author[3]{Aaron J. Rosengren}
\credit{Conceptualization, Supervision, Formal analysis, Investigation, Visualization, Writing - original draft, Writing - review \& editing}

\affiliation[inst3]{organization={Mechanical and Aerospace Engineering, UC San Diego},
            addressline={9500 Gilman Drive}, 
            city={La Jolla},
            postcode={92093}, 
            state={CA},
            country={USA}}

\cortext[cor1]{Corresponding author}


\begin{abstract}
  We present a general analytic framework to assess whether impact ejecta launched from the surface of a satellite can escape the gravitational influence of the planet--satellite system and enter heliocentric orbit. Using a patched-conic approach and defining the transition to planetocentric space via the Hill sphere or sphere of influence, we derive thresholds for escape in terms of the satellite-to-planet mass ratio and the ratio of the satellite’s orbital speed to its escape speed. We identify three dynamical regimes for ejecta based on residual speed and launch direction. We complement this analysis with the circular restricted three-body problem, deriving a necessary escape condition from the Jacobi integral at $\mathrm{L_{2}}$ and showing that it is consistent with the patched–conic thresholds. Applying our model to the Earth--Moon system reveals that all three outcomes---bound, conditional, and unbound---are accessible within a narrow range of launch speeds. This behavior is not found in other planetary satellite systems, but may occur in some binary asteroids. The framework also shows that the Moon's tidal migration has not altered its propensity to produce escaping ejecta, reinforcing the plausibility of a lunar origin for some near-Earth asteroids.

\nocite{*}

\end{abstract}


\begin{highlights}

\item Simple formulae estimate launch speed of satellite ejecta to reach heliocentric space

\item Escape to heliocentric space depends on mass ratio and orbital-to-escape-speed ratio

\item Earth--Moon lies uniquely close to the critical condition for barely escaping ejecta

\item Moon’s migration has little effect on the fates of ejecta 

\end{highlights}


\maketitle


\section{Introduction}
\label{sec:Introduction}

The study of meteoroidal ejecta---fragments produced  by high-velocity impacts on the surfaces of natural satellites---offers critical insights into a wide range of planetary processes.
These include crater formation, regolith redistribution, secondary cratering, material exchange among planetary satellites, and the delivery of debris into interplanetary space \citep{Soderblom1970,Melosh_1989,Gladman1996,McEwen2006}.
Pioneering work by \citet{Gladman1995} on the dynamical evolution of lunar ejecta demonstrated that a 
significant
fraction of particles launched from the Moon at near-escape speeds
fall onto Earth within a few tens of thousands of years, while a smaller fraction escapes the Earth--Moon system on heliocentric trajectories, with some eventually colliding with other planets.
Similarly detailed studies by \citet{Alvarellos2005,Alvarellos2017} investigated the fates of ejecta from satellites in the Saturnian system, revealing a rich array of outcomes including re-accretion by the source satellite, collisions with other satellites, and a much smaller portion escaping into heliocentric orbits. 
\citet{Dobrovolskis2010} also showed that ejecta from Calypso and Telesto  in the Saturnian system can reach co-orbital configurations around Tethys.

Recent discoveries have reinvigorated interest in the topic of the fates of lunar ejecta.
The near-Earth asteroid (469219) Kamo'oalewa has been suggested to originate from the Moon, owing to its L-type reflectance spectrum and Earth-like orbit \citep{Sharkey2021}. 
Numerical simulations suggest that this lunar origin is compatible with its dynamics, in particular, that a lunar ejecta particle can reach 
a stable co-orbital configuration with the Earth's heliocentric orbit \citep{Castro2023,Castro2025a,Jiao2024}. 
Additionally, the newly identified object 2024 PT$_5$ has orbital properties consistent with a possible lunar or Earth--Moon system origin \citep{Kareta2025}.
In related work, we found that the likelihood of lunar ejecta impacting Earth is
highly sensitive to the launch location on the Moon's surface, showing a pronounced asymmetry between the trailing and leading lunar hemispheres, with ejecta from the trailing side being significantly more likely to reach Earth \citep{Castro2025b}.
These findings prompt renewed inquiry into the role of lunar ejecta fragments in shaping the near-Earth object (NEO) population. 

Previous efforts to track the trajectories of satellite ejecta have relied primarily on partial or full N-body numerical integrations over long time scales. 
While these simulations provide high-fidelity calculations of their pathways, they may not easily reveal the underlying conditions that govern the fate of the ejecta particles. 
In a previous work \citep{Castro2023}, we used a patched-conics approximation to study the fate of lunar ejecta, deriving simple conditions for a particle to escape both the Moon and Earth and enter heliocentric orbit. That analysis revealed the seemingly coincidental result that particles launched near the escape speed from the Moon’s leading side are just able to escape to heliocentric orbit. 
While this finding has important implications for the efficiency of the Moon as a source of escaping material, it also raised a question on the generality of this phenomenon and the conditions leading
to it. 
Building on the analysis of \citet{Castro2023}---itself reminiscent of the earlier work by \citet{Shute1966}---this paper generalizes the framework to satellite–planet systems beyond the Earth--Moon case.
We develop a simplified but broadly applicable formalism
to estimate the fate of ejecta launched from the surface of natural satellites and derive analytic thresholds for ejecta particles to escape the gravitational influence of the planet--satellite. While these calculations are approximate, they offer a general picture of the fates of a satellite's ejecta fragments based on simple criteria that depend on only a few parameters, such as the mass ratio of the satellite and its host planet and the ratio of the satellite's surface escape velocity to its orbital velocity, which allow to identify general trends and assess how unusual the Earth--Moon configuration truly is when compared with other satellites.

The structure of this paper is as follows. 
In Section~\ref{sec:Theory}, we derive theoretical estimates for the minimum ejection speeds necessary for ejecta particles to escape into interplanetary space when launched from different locations on the surface of the satellite. We then employ the circular, restricted three-body problem (CRTBP) framework to derive a necessary condition based on the value of the Jacobi integral at $\mathrm{L_{2}}$. This CRTBP constraint is consistent with the escape-speed estimates based on the patched-conic approximation. 
In Section~\ref{sec:Discussion}, we discuss the implications of these results for satellite--planet pairs in the Solar System, paying particular attention to Earth--Moon system (including early lunar tidal migration). We summarize our results in Section~\ref{sec:Summary}.

\section{Theoretical calculations}
\label{sec:Theory}

\subsection{Estimates from the method of patched conics}
\label{subsec:patched}

Consider a satellite of mass $m_{\text{S}}$ and radius $R_{\text{S}}$ on a circular orbit at a distance $a_{\text{S}}$ from a planet of mass $m_{\text{P}} \gg m_{\text{S}}$. 
For later use, we define the satellite-to-planet mass ratio, $\mu$,
\begin{equation}
  \mu = \frac{m_\text{S}}{m_\text{P}}.
  \label{mu}
\end{equation}
We also note that the satellite's orbital speed around the planet is given by
\begin{equation}
  v_{\text{orb,S}} = \sqrt{\frac{G(m_\text{P}+m_\text{S})}{a_{\text{S}}}},
  \label{orb_S}
\end{equation}
and the satellite's escape speed is given by
\begin{equation}
  v_{\text{esc,S}} = \sqrt{\frac{2Gm_{\text{S}}}{R_{\text{S}}}},
  \label{escape_S}
\end{equation}
where $G$ is the universal constant of gravitation. 
Table~\ref{table1} lists the values of $\mu$ and of the ratio, $v_\mathrm{orb,S}/v_\mathrm{esc,S}$, for a selection of planet--satellite pairs in the Solar System; in each case, the satellite listed is the largest satellite of its host planet.
We can observe that our assumption regarding $\mu\ll 1$ is a good approximation in most cases. 
The largest mass ratio in Table~\ref{table1} is that of Charon-to-Pluto, with $\mu \approx 0.12$; in this case we may expect the patched-conic estimates to be less reliable than in the other cases.

We analyze the motion of a particle of negligible mass launched from the surface of the satellite with speed $v_{\text{L}}$ and focus on particles launched with a speed $v_{\text{L}}$ exceeding the satellite's escape speed; i.e., $v_{\text{L}} \ge v_{\text{esc,S}}$.
In the patched-conics framework, we define a spherical boundary of radius $r_\mathrm{B}$ centered at the satellite. Within this sphere, we approximate the particle's motion as a conic-section trajectory under the gravity of only the satellite. Outside of this sphere, we approximate the particle's motion as a conic-section trajectory under the central gravity of a point mass, $m_\mathrm{P}+m_\mathrm{S}$, located at the planet--satellite barycenter.
For the following estimates, we will take $r_\mathrm{B}$ to be the satellite's Hill-sphere radius. 
That is, we define the boundary between the satellite-centric and planetocentric space as the satellite's Hill sphere, whose radius is calculated as
\begin{equation}
  r_{H,\text{S}} = a_\text{S} \left( \frac{\mu}{3(1+\mu)} \right)^{1/3}.
  \label{hill}
\end{equation}
A particle launched from the surface of the satellite with $v_{\text{L}} \geq v_{\text{esc,S}}$ will reach the satellite's Hill-sphere boundary with a residual speed $\delta v_{\text{L}}$, as given approximately by the conservation of energy in the satellite's gravitational field,
\begin{equation}
	\delta v_{\text{L}}^2 
	 = v_{\text{L}}^2 - \frac{2 G m_\text{S}}{R_\text{S}} \left( 1 - \frac{R_\text{S}}{r_{{H},\text{S}}} \right).
        \label{residual}
\end{equation}
We next examine whether particles reaching the Hill sphere with this residual speed  can escape into interplanetary space. 
Here, `escape' will be used to refer to a particle with sufficient speed to be unbounded from the planet--satellite system and enter a heliocentric orbit. 
The velocity of a launched particle relative to the planet--satellite barycenter, 
$\mathbf{v_{\text{B}}}$, is the vector sum of  the residual velocity at its Hill sphere 
and the satellite's  barycentric orbital velocity, which is approximately given by $(1-\mu) \mathbf{v}_\mathrm{orb,S}$: 
\begin{equation}
  \mathbf{v_{\text{B}}} = \mathbf{\delta v_{\text{L}}} + (1-\mu)\mathbf{v_{\text{orb,S}}}.
  \label{barycenter}
\end{equation}
To escape  the planet--satellite system, the barycentric speed must exceed the local barycentric escape speed. 
The local barycentric speed depends on the particle's position when it reaches the Hill sphere, which is determined by the launch conditions. 
For simplicity, we  rely on the assumption that $\mu$ is small, so the Hill sphere is small compared with the satellite's orbital radius. Under this approximation, the local escape speed can be estimated as the escape speed at the satellite's barycentric distance, given by 
$\sqrt2(1-\mu) v_\mathrm{orb,S}$.

\begin{table}[h]
\centering
\begin{tabular}{l l c c }
\hline
\textbf{Planet} & \textbf{Satellite} & $\mu$ & $v_{\rm orb,S}/v_{\rm esc,S}$  \\
\hline
Earth   & Moon     & $1.23 \times 10^{-2}$ & $0.43$\\
Mars    & Phobos   & $1.66 \times 10^{-8}$ & $190$\\
Jupiter & Ganymede & $7.81 \times 10^{-5}$ & $3.97$\\
Saturn  & Titan    & $2.37 \times 10^{-4}$ & $2.1$\\
Uranus  & Titania  & $4.06 \times 10^{-5}$ & $4.7$ \\
Neptune & Triton   & $2.09 \times 10^{-4}$ & $3.0$ \\
Pluto   & Charon   & $1.22 \times 10^{-1}$  & $0.35$ \\

\hline
\end{tabular}
\caption{Parameters of the largest natural satellite of each planet in the Solar System. The third column is the ratio of the satellite to planet mass, $\mu$, and the fourth column is the ratio of the escape speed from the satellite's surface to its planetocentric orbital speed.}
\label{table1}
\end{table}

The result of Eq.~(\ref{barycenter}) depends on the direction and location of launch. The two bounding cases are vertical launches from the satellite's trailing and leading hemispheres.
From the leading side, the particle gets a positive boost aligned with the satellite's orbital velocity and it will therefore have the largest barycentric speed; conversely, from the trailing side,
the particle's velocity is anti-aligned with the satellite's orbital velocity and will therefore have the smallest barycentric speed. For these bounding cases, escape requires
\begin{equation}
    \delta v_{\text{L}} \geq (\sqrt{2}(1-\mu)\pm1) v_{\mathrm{orb},\text{S}} \approx (\sqrt2\pm 1) v_{\mathrm{orb},\text{S}}. 
    \label{bounds}
\end{equation}
Any other launch scenario will lead to a residual speed that lies between these two limiting cases.  
The bounds from Eq.~(\ref{bounds}) define three dynamical regimes: 
(i) if $\delta v_{\text{L}} < (\sqrt{2}-1) v_{\mathrm{orb},\text{S}}$ no particles escape; 
(ii) for $(\sqrt{2}-1) v_{\mathrm{orb},\text{S}} <\delta v_{\text{L}} < (\sqrt{2}+1) v_{\mathrm{orb},\text{S}}$  escape is possible if the direction of the residual velocity is sufficiently close to the direction of the satellite's velocity; 
(iii) if $\delta v_{\text{L}} > (\sqrt{2}+1) v_{\mathrm{orb},\text{S}}$, escape occurs for any launch geometry. 

\begin{figure}
    \centering
    \includegraphics[width=0.9\textwidth]{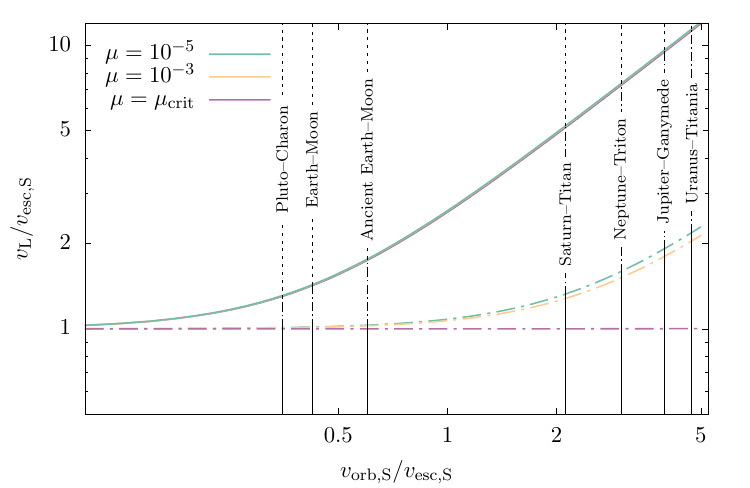}   
    \caption{The threshold launch velocities required for escape to heliocentric orbit.
Curves in different colors represent different cases of the satellite-to-planet mass ratio, $\mu$ (\textit{purple}, \textit{yellow} and \textit{green} for $\mu_\mathrm{crit}=0.0147$, $10^{-3}$, and $10^{-5}$, respectively). The \textit{dashed} curves represent the minimum launch speed to escape with a vertical launch from the leading side, and the \textit{solid} curves represent the minimum launch speed to escape with a vertical launch from the trailing side. 
Launch velocities above the \textit{solid} curves always escape, whereas launch velocities below the \textit{dashed} curves do not; launch velocities in-between the \textit{solid} and \textit{dashed} curves may escape to heliocentric orbit depending upon launch location and direction of launch. 
The vertical lines indicate parameter values of several satellite--planet pairs. } 
    \label{fig:dv}
\end{figure}

Substituting Eq.~(\ref{bounds}) into Eq.~(\ref{residual}), we find that the launch speed needed to escape for each of the two bounding cases is given by 
\begin{equation}
    v_{\text{L}}^{2} \geq  \frac{2Gm_{S}}{R_{S}} \left( 1- \frac{R_{S}}{r_{H,S}} \right)
    + \left( \sqrt{2} \pm 1 \right)^{2} v_{\text{orb,S}}^{2}.
\label{e:vLsq}
\end{equation}
We can express the ratio of the satellite's and Hill's radii as 
\begin{equation}
    \frac{R_{S}}{r_{H,S}} = \frac{R_{S}}{G m_{S}} \frac{G (m_{P}+m_{S})}{a_{S}} 
    \frac{m_{S}}{m_{P}+m_{S}} 
    \left( \frac{3 (m_{P}+m_{S})}{m_{S}} \right)^{1/3} 
    = f(\mu) \frac{v_{\text{orb,S}}^{2}}{v_{\text{esc,S}}^{2}},
\label{e:Rratio}
\end{equation}
where $f(\mu)$ is a positive function of $\mu$,
\begin{equation}
f(\mu)  = 2 \sqrt[3]{3}  \Bigl(\tfrac{\mu}{1+\mu}\Bigr)^{2/3}.
\label{f}
\end{equation}
Substituting Eq.~\eqref{e:Rratio} into Eq.~\eqref{e:vLsq} and re-arranging, we can cast the escape condition as follows:
\begin{equation}
\left(\frac{v_{\text{L}}}{v_{\text{esc,S}}}\right)^2 \;\ge\; 1 \;+\;
\Bigl[(\sqrt2\pm1)^2 - f(\mu)\Bigr]
\left( \frac{v_{\rm orb,S}}{v_{\rm esc,S}} \right)^2.
\label{vl}
\end{equation}
We observe that this condition depends on two parameters that characterize the satellite--planet system: 
the mass ratio $\mu$ and the ratio of the orbital and escape speed of the satellite $v_{\text{orb,S}}/v_{\text{esc,S}}$. 
In Fig.~(\ref{fig:dv}) we plot the threshold values of $v_{\text{L}}/v_{\text{esc,S}}$ as a function of $v_{\text{orb,S}}/v_{\text{esc,S}}$, for three different values of $\mu$.
The dashed and solid curves correspond to leading-side and trailing-side launches, respectively. 
Particles launched with speeds between the curves may escape, depending on the geometry of launch; 
those above the curves always escape. 

For small close-in satellites, the minimum launch speeds for escape can be significantly larger than the surface escape speed of the satellite.
However, we also observe the interesting property that the minimum launch speed for escape is as small as the satellite's escape speed when the second term in Eq.~(\ref{vl}) vanishes. 
This occurs when
\begin{equation}
f(\mu)= (\sqrt2 \pm 1)^{2} \qquad\hbox{for}\ \mu=\mu_\mathrm{crit}.
\label{fmu_crit}
\end{equation}
Only the minus sign gives a physically meaningful solution ($\mu<1$):
\begin{equation}
\frac{\mu_\mathrm{crit}}{1+\mu_\mathrm{crit}} = \Bigl[\frac{(\sqrt2-1)^{2}}{2\sqrt[3]{3}}\Bigr]^{3/2}
\quad\Longrightarrow\quad
\mu_\mathrm{crit}\approx 0.0147. 
\label{mu_crit}
\end{equation}
For this value, ejecta launched barely above escape speed from the satellite’s leading side can enter heliocentric orbit. Note that since Eq.~(\ref{fmu_crit})  does not have a physical solution for the positive branch, there is no value of $\mu$ that guarantees escape for an arbitrary launch  and $v_\text{orb,S}/v_{\text{esc,S}}$ value. \\

\begin{figure}
    \centering
    \includegraphics[width=0.9\textwidth]{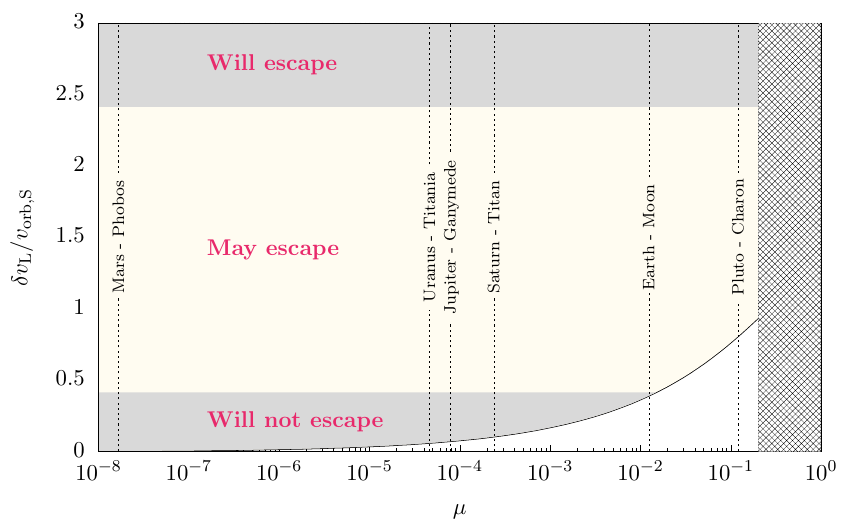}   
    \caption{Ratio of residual speed at the satellite's Hill sphere and the satellite orbital speed,  versus the satellite-to-planet mass ratio $ \mu $.  The curve shows the bounding case of particles launched with speed equal to the satellite's escape velocity. \textit{Shaded} regions represent the regions of different outcomes for ejecta launched above escape velocity. The \textit{stippled} region is the range of $\mu>0.2$ where the patched-conics approximation becomes less effective.} 
    \label{fig:dv2}
\end{figure}

From Eq.~(\ref{residual}), we see that any particle with $v_{\text{L}} \geq v_{\text{esc,S}}$ will reach the satellite's Hill sphere with a residual speed given by
\begin{equation}
    \delta v_{\mathrm{L}}^{2} \geq \frac{R_{S}}{r_{H,S}} v_{\text{esc,S}}^{2} = f(\mu) v_{\text{orb,S}}^{2}. 
    \label{lower_dv}
\end{equation}
Together with the bounds from Eq.~(\ref{bounds}), we can define the three dynamical regimes based only on the satellite-to-planet mass ratio, $\mu$. 
This is illustrated in Fig.~(\ref{fig:dv2}), where the residual speed of ejected particles relative to the orbital speed can have values above the solid black curve defined by the bounds of Eq.~(\ref{lower_dv}). 
Notice how the bound from Eq.~(\ref{lower_dv}) meets the lower bound from Eq.~(\ref{bounds}) at $\mu=\mu_\mathrm{crit}$. 
It is striking that the Earth--Moon $\mu$ value is close to $\mu_\mathrm{crit}$; 
consequently, there is only a very narrow range of lunar-launch velocities that would not result in escape to interplanetary space. 

\subsection{Estimates from the circular, restricted three-body problem (CRTBP)}
\label{subsec:CRTBP}

To complement the patched-conic estimates from the previous subsection, we next employ the 
CRTBP to assess conditions for heliocentric escape. For use in the following calculations, we define the satellite-to-system mass ratio, $\bar{\mu}$, as
\begin{equation}
    \bar{\mu} = \frac{m_{\mathrm{S}}}{m_{\mathrm{P}}+\mathrm{m_{S}}}.
\end{equation}
In the CRTBP framework, the Jacobi integral $C_{\mathrm{J}}$ delineates the regions of allowed and forbidden motion and is expressed, in the rotating (synodic) frame, in terms of the particle's position and velocity as
\begin{equation}
    C_{\mathrm{J}} = 2U - v^{2},
\end{equation}
where the pseudo-potential $U$ is defined as 
\begin{equation}
    U = \frac{1}{2} \left(x^{2} +y^{2} \right) + \frac{1-\bar{\mu}}{r_{\mathrm{P}}}+\frac{\bar{\mu}}{r_{\mathrm{S}}},
\end{equation}
and the distances to the planet and satellite are
\begin{equation}
    r_{\mathrm{P}}^{2} = \left(x+\bar{\mu} \right)^{2} + y^{2} + z^{2},
\end{equation}
\begin{equation}
    r_{\mathrm{S}}^{2} = \left(x-1+\bar{\mu} \right)^{2} + y^{2} + z^{2}.
\end{equation}
Here we adopt the units in which the planet--satellite distance $a_{\mathrm{S}}=1$ and
$G(m_{\mathrm{P}}+m_{\mathrm{S}})=1$, so that the satellite's orbital speed is $v_{\mathrm{orb,S}}=1$.   

\begin{figure}[]
	\begin{center}
	\includegraphics[width=0.495\textwidth]{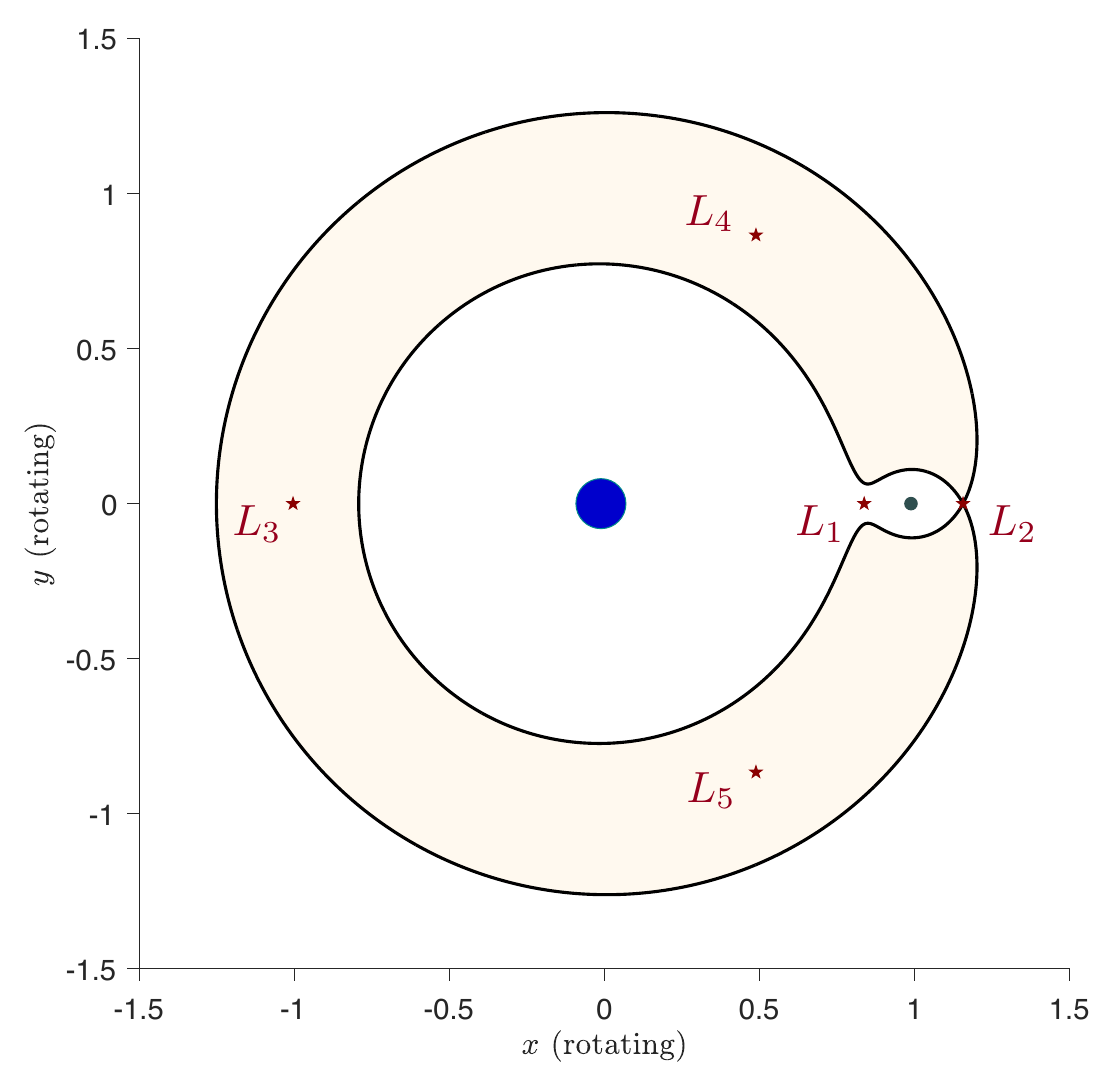}
	\includegraphics[width=0.495\textwidth]{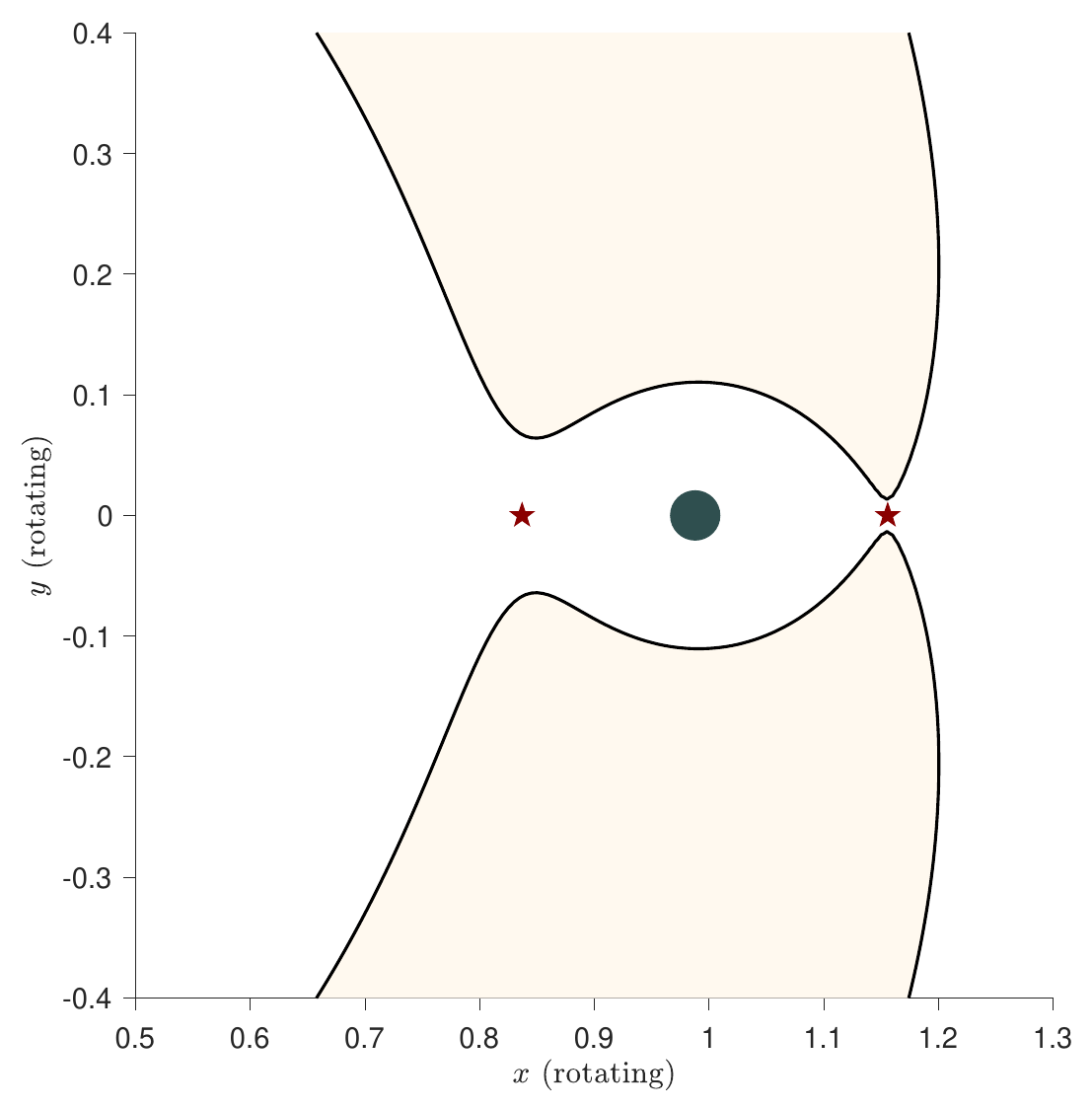}
	\caption{
	Forbidden regions ({\itshape shaded}) within the Earth--Moon system for $C_\mathrm{J} = C_{\mathrm{L_{2}}} $ ({\itshape left}) and the location of the Lagrange equilibrium points (denoted by {\itshape five-pointed stars}). Decreasing the Jacobi integral below $C_{\mathrm{L_2}}$ ({\it right}) opens the corridor or bottleneck around $\mathrm{L_2}$ enabling escape from the Earth--Moon system. 
	}
	\label{fig:zvc}
	\end{center}
\end{figure}

Since the particle's speed is non-negative, it follows that $C_{\mathrm{J}} \leq 2U$, with
equality defining a boundary between forbidden and allowed regions of motion. In the
planar case ($z=0$, $\dot{z}=0$), this condition produces a set of zero-velocity curves. When the Jacobi integral is sufficiently large, the region of allowed motion is disconnected into two or even three disjoint parts, and the motion remains confined to whichever zone it initially occupies. 
Fig.~(\ref{fig:zvc}) shows the Jacobi integral at the equilibrium Lagrange point $\mathrm{L_{2}}$, denoted $C_{\mathrm{L_{2}}}$, for the Earth--Moon system. At this energy level, the three realms of possible motion are interconnected, and escape is enabled through a narrow corridor in the vicinity of $\mathrm{L_{2}}$. The escape criteria can therefore be written as
\begin{equation}
C_{\mathrm{J}}=2U-v^{2}\leq C_{\mathrm{L_{2}}},
\label{jacobi_escape}
\end{equation}
providing a necessary condition on the launch speed for which escape becomes possible. 
For small $\bar{\mu}$, the Jacobi integral at $\mathrm{L_{2}}$ can be approximated as \citep{cMsD99}
\begin{equation}
    C_{\mathrm{L_{2}}} \approx 3 + 3^{4/3}\bar{\mu}^{2/3}- \frac{10}{3} \bar{\mu}.
    \label{cl2}
\end{equation}
The pseudo-potential at the surface of a small satellite varies only slightly with launch location, and its value can be approximated  by evaluating at $x=1-\bar{\mu}$, $y=\pm R_{\mathrm{S}}$ $z=0$
\begin{equation}
   2 U_{\mathrm{S}} = (1-\bar{\mu})^{2} + \frac{2(1-\bar{\mu})}{\sqrt{1+R^{2}_{\mathrm{S}}}} + \frac{2\bar{\mu}}{R_{S}} \approx 3 - 4\bar{\mu} +\frac{2\bar{\mu}}{R_{S}}.
   \label{potential}
\end{equation}
Now we consider a particle launched from the satellite’s surface with speed $v_{\mathrm{L}}$.
While $v_{\mathrm{L}}$ is measured in a topocentric frame at the surface of the satellite, it is the same as the speed $v$ in the synodic frame appearing in Eq.~\eqref{jacobi_escape}, provided the satellite is stationary in that frame. 
When the center of mass is fixed, all surface points remain stationary if the satellite is tidally locked with its host---as is the case for the large regular satellites of the Solar System, such as Ganymede, Titan, and Charon; in other cases the analysis below neglects the effects of non-synchronous rotation of the satellite and $v_{\mathrm{L}}$ refers to the launch speed in the synodic frame.Substituting Eqs.~(\ref{cl2}) and (\ref{potential}) into the escape condition (\ref{jacobi_escape}) yields
\begin{equation}
    v^{2}_{\mathrm{L}} \geq \frac{2\bar{\mu}}{R_{\mathrm{S}}} -3^{4/3} \bar{\mu}^{2/3} + \frac{2}{3}\bar{\mu}. 
\end{equation}
Transforming back to dimensional units, we apply $v_{\mathrm{L}} \rightarrow v_{\mathrm{L}}/v_{\mathrm{orb,S}}$ and $R_{\mathrm{S}} \rightarrow R_{\mathrm{S}}/a_{\mathrm{S}}$. Using Eqs.~(\ref{orb_S}) and (\ref{escape_S}) the escape condition becomes
\begin{equation}
\left(\frac{v_{\text{L}}}{v_{\text{esc,S}}}\right)^2 \;\ge\; 1 \;-\;
\Bigl[ 3^{4/3} \bar{\mu}^{2/3} - \frac{2}{3} \bar{\mu}  \Bigr]
\left(\frac{v_{\rm orb,S}}{v_{\rm esc,S}}\right)^2.
\label{jacobi_vl}
\end{equation}
We note that the lower bound in Eq.~(\ref{jacobi_vl}) is always smaller than the satellite's surface escape speed, whereas the lower bound derived in Subsection~(\ref{subsec:patched}) (Eq.~\ref{vl}) is always at least equal to it. This difference comes from the fact
that the CRTBP-based condition in Eq.~(\ref{jacobi_escape}) is less restrictive; it requires only that the particle be able to access the exterior region beyond the $\mathrm{L_{2}}$ bottleneck, but not necessarily achieve a heliocentric orbit. In contrast, Eq.~(\ref{bounds}) imposes the stronger condition that the particle become unbound from the gravitational influence of the planet and satellite.

\section{Discussion} 
\label{sec:Discussion}
\subsection{Applications to Solar System planetary satellites}

The theoretical framework developed in Section~\ref{sec:Theory} provides approximate analytic criteria for how ejecta launched from the surface of a satellite can escape to heliocentric orbit.
We list in Table~\ref{table1} 
the pertinent parameters for these criteria for the largest planetary satellites in the Solar System. 
Among these, the regular moons of the giant planets have small $\mu$ values (typically $ \mu \lesssim 10^{-4} $), well below the threshold identified in Eq.~(\ref{mu_crit}) and low $ v_{\text{orb,S}}/v_{\text{esc,S}}$  ratios (of the order of unity). 
For these systems, ejecta launched near the escape speed are unlikely to escape to heliocentric space, and the launch speed threshold to escape spans over several times the satellite escape speed, as shown in Fig.~(\ref{fig:dv}). 

We compare our analytic estimates with the specific case of Jupiter’s moon Ganymede, for which previous studies have reported numerical results on the fates of launched ejecta.
Our analytic estimates predict that  launch speeds below 1.9 times Ganymede's escape speed would be unlikely to result in heliocentric escapes. 
This expectation is consistent with the numerical results of \citet{Alvarellos2002} who performed N-body simulations of ejecta from the Gilgamesh basin on Ganymede using launch speeds between 0.90 and 1.4 times the  escape speed. 
They found that only a small fraction of particles (1–3.6\%, depending on launch direction) ultimately escaped to heliocentric space. 
All the cases of heliocentric escape occurred after more than $\sim 1000$ years.
The fact that escape required more than a millennium in their simulations suggests that more complex gravitational perturbations--from multiple Jovian moons and the Sun and Saturn--had time to accumulate, resulting in these rare escape cases.

In contrast, the irregular moons of the giant planets often have larger semi-major axes and lower escape speeds. 
While their $\mu$ is typically smaller, their \( v_{\text{orb,S}}/v_{\text{esc,S}} \) ratios can be significantly higher due to their distant orbits. For instance, for the irregular Jovian satellite Themisto, \( v_{\text{orb,S}}/v_{\text{esc,S}} \)~$\sim 1200$, making the entrance to heliocentric space of ejecta highly prohibitive, requiring to launch with a speed at least $\sim 500$ larger than the surface escape speed. 
Something similar happens in the case of Mars's satellites, given their small size and relatively distant orbits. 

On the opposite side is the case of Pluto-Charon. 
For this system, $\mu=0.122$, which is above the critical value calculated in Eq.~(\ref{mu_crit}). 
Although this not-so-small mass ratio means that our analytic estimates may have limited accuracy, it is still interesting to compare these estimates with numerical results in the literature. 
Our analytic estimates based on the patched conics framework predict that heliocentric escape would be nearly guaranteed for ejecta launched from Charon's leading side with velocity above Charon's escape velocity of 0.58~km~s$^{-1}$ and for ejecta launched with velocity exceeding 0.73~km~s$^{-1}$ from any location on Charon.
Our estimate based on the CRTB model (Eq.~\eqref{jacobi_vl}) predicts a minimum launch speed of $\sim 0.51$~km~s$^{-1}$ to allow heliocentric escape (however, as we discussed previously, this is a necessary but not sufficient condition).
The results of N-body simulations by \citet{Bierhaus2015} are consistent with our analytic estimates. 
In these simulations, the authors included perturbations from all known moons in the Pluto system, and sampled Charon ejecta launch speeds in the range 0.55--1~km~s$^{-1}$. 
They found that the minimum speed for Charon ejecta to escape to heliocentric orbit is 0.55~km~s$^{-1}$.
They also reported heliocentric escape for most ejecta launched with speeds above 0.69~km~s$^{-1}$.  
The agreement with our analytic approximations is notable despite the not-so-small mass ratio of Charon-to-Pluto.

We also note an interesting contrast between Ganymede and Charon when comparing the necessary and sufficient conditions derived from our two approaches. For Ganymede, the minimum necessary speed for heliocentric escape from the CRTB model is $0.88 v_{\mathrm{esc,S}}$, while the minimum sufficient speed from the patched-conics estimate is $1.9 v_{\mathrm{esc,S}}$, more than twice as large. This wide separation implies that particles launched at speeds near or below the satellite’s escape velocity are very unlikely to reach heliocentric space. In contrast, for Pluto–Charon, the necessary and sufficient conditions are $0.89 v_{\mathrm{esc,S}}$ and $v_{\mathrm{esc,S}}$, respectively. In this case, particles launched only marginally below Charon’s escape speed already satisfy the necessary condition, making heliocentric escape more likely, and our sufficient condition may overestimate slightly the actual minimum speed for escape.

\subsection{The unique Earth--Moon system}

The Earth--Moon system hosts unique dynamical behaviors among known planet--satellite systems in the Solar System. 
As shown in Section~\ref{sec:Theory}, the conditions for escape of ejecta from a satellite into heliocentric orbit depends on two parameters: the satellite-to-planet mass ratio $\mu$, and the ratio of the satellite's orbital velocity to its escape velocity, $v_{\mathrm{orb},S} / v_{\mathrm{esc},S}$. 
For the Moon, both parameters conspire to reduce the launch speed needed for such escape. 

Specifically, the Moon's mass ratio $\mu = 0.0123$ lies just below the critical value 
$\mu_{\mathrm{crit}} \approx 0.0147$ derived in Eq.~(\ref{mu_crit}). 
This proximity ensures that ejecta launched at barely above the lunar escape speed can reach the Earth--Moon Hill sphere with residual speeds near the minimum needed for escape. 
Furthermore, the Moon's orbital speed is relatively small compared to its surface escape speed: $v_{\mathrm{orb,S}}/v_{\mathrm{esc},S} \approx 0.43$. 
As a result, the range of launch speeds for conditional escape is narrow: from just 1.002 times $v_{\mathrm{esc},S}$ (for favorable leading-side launches) to 1.43 times $v_{\mathrm{esc},S}$ (for unfavorable trailing-side launches). 
This threshold delineates all three possible dynamical outcomes: (i) remaining bound to the Earth--Moon system, (ii) conditional escape, and (iii) guaranteed escape to interplanetary space.

No other planet--satellite system in the Solar System exhibits this combination of mass ratio and speed ratio. Most systems have much smaller $\mu$, resulting in higher launch speeds required for heliocentric escape (particularly for launch from their trailing side).

The results applied to the Earth–Moon case are consistent with our previous numerical studies. In \citet{Castro2023}, we found that the ability of a particle to become a co-orbital depends sensitively on both its launch speed and its ejection site on the Moon. In particular, ejecta launched from the trailing hemisphere required speeds exceeding approximately $3.4 \, \mathrm{km \, s^{-1}}$ to reach Earth's co-orbital space, whereas particles launched from the leading hemisphere could enter such orbits with speeds just above the lunar escape speed.
\citet{Castro2025b} further confirmed this asymmetry for particles colliding with Earth, showing that those launched from the trailing side are more likely to strike the planet, either by remaining bound to the Earth–Moon system or by achieving very Earth-like heliocentric orbits. In contrast, particles launched from the leading side were least likely to hit Earth and more likely to escape into non-colliding heliocentric trajectories.

\subsection{Moon's outward migration}

An important feature of the Earth--Moon system is the Moon's tidal migration over geological timescales. 
This process, driven by angular momentum transfer from Earth’s rotation, has secularly increased the orbital radius of the Moon since its formation. 
Geological and dynamical evidence suggests that the Moon was once significantly closer to Earth, with a correspondingly higher orbital velocity \citep{Meyer2010}.

Crucially, the satellite-to-planet mass ratio $\mu$ remains constant during this evolution. 
The dynamical parameter that changes is the ratio $v_{\mathrm{orb},S}/v_{\mathrm{esc},S}$, which decreases as the Moon's orbital radius increases. 
Since the escape speed from the Moon’s surface is independent of its orbital distance, the decreasing orbital speed leads to a smaller value of $v_{\mathrm{orb},S}/v_{\mathrm{esc},S}$.

Despite this evolution, the Earth--Moon system remains close to the critical escape threshold throughout its history. For instance, even if the Moon's orbital radius were halved from its present-day value, the corresponding speed ratio would increase to only about $0.6$. 
This is still a sufficiently small value, implying a narrow range of launch speeds sufficient for conditional escape. 
Therefore, the conclusion that lunar ejecta can access all three dynamical fates remains valid throughout much of the Moon’s tidal migration.

This has implications for the early Solar System. 
During past epochs of intense bombardment, such as the Moon's formation and the Late Heavy Bombardment \citep{Bottke2017}, the Moon's orbital distance was likely smaller, yet the same dynamical sensitivity to ejecta speed and direction would have persisted. Over geological timescales, the continuous transfer of lunar material into near-Earth space could have contributed not only to Earth’s impact history but also to the exchange of surface material between the two bodies, potentially contributing to the observed isotopic similarities between Earth and Moon samples \citep{Nielsen2021}.

The persistence of favorable dynamical conditions throughout the Moon’s orbital evolution indicates that ejecta from large impacts could escape the Earth–Moon system and populate Earth’s co-orbital space in the past. 
More recent events, such as the formation of Giordano Bruno impact crater 1–10 Myr ago \citep{Jiao2024}, could have produced ejecta that evolved into contemporary quasi-satellite configurations like that of Kamo‘oalewa \citep{Castro2023}. 
The orbital stability timescale of such objects is estimated to be a few Myr \citep{dFM2016,Fenucci2021,Liu2022}, sufficiently long for fragments from relatively recent large impacts to survive until the present day.

\subsection{Binary asteroids}

A dynamical analogue of escaping particles from a planet--satellite system is a binary asteroid system. 
The range of mass ratios for these systems extends from $\mu \sim$$10^{-3}$ to values near unity. 
Thus, we can apply the results from Section~\ref{sec:Theory} for a subset of these asteroids with a small $ \mu \lesssim 10^{-1} $. 
One particularly interesting example is the Didymos--Dimorphos system, where the estimated mass ratio is $\mu \approx 8.3 \times 10^{-3}$. 
While this value is below but relatively close to the critical threshold, the orbital-to-escape speed ratio for Dimorphos is much larger than in the Earth--Moon case ($v_{\mathrm{orb,S}}/v_{\mathrm{esc,S}} \approx 12.2$). 
Substituting into Eq.~(\ref{vl}) results in a much broader range of required launch speeds for heliocentric escape: between 3.05 and 29 times $v_{\mathrm{esc,S}}$, depending on launch geometry. 
Thus, although Dimorphos shares a similar mass ratio with the Earth--Moon, its high ratio of orbital-to-escape speed makes heliocentric escape far more difficult. 

\subsection{Caveats}

The analytic derivations presented here are based on simplified assumptions, as noted. 
In particular, we assumed small values for the satellite-to-planet mass ratio in order to obtain simple formulas for heliocentric escape as derived in Eq.~(\ref{bounds}). 
While this assumption is reasonable for most satellite-to-planet pairs, it would not be
adequate for studying binary asteroids systems with similar-sized components. Below we list a number of other limitations of our formulas.

\begin{itemize}

\item Our derivations  in Section~\ref{sec:Theory}  use the Hill sphere as the transition boundary between satellite-dominated space to planet-dominated space. 
However, this boundary should be regarded as an approximate construct, since gravitational influence is continuous and long-ranged, rendering the transition between satellite- and planet-dominated regimes inherently diffuse.
An alternative definition of the boundary is the sphere of influence \citep{gC64}, defined as   
\begin{equation}
  r_{\text{SOI}} = a \left( \frac{m_{S}}{m_{P}} \right)^{2/5}.
\end{equation}
This alternative choice leads to a slightly different form of the function $f(\mu)$ we derived in Eq.~(\ref{f}), 
\begin{equation}
    \widetilde f(\mu) = \frac{2\mu^{3/5}}{1+\mu},
    \label{e:f_soi}
\end{equation}
which yields a slightly different critical value $\widetilde\mu_{\text{crit}}=0.0172$. 
The differences between $f(\mu)$ and $\widetilde f(\mu)$ are significant for larger values of $\mu$, and are at most $\sim$$30\%$ for the range of satellite-to-planet mass ratios found in the Solar System. 
We note that for mass ratios closer to 1 (approximately larger than $4\times 10^{-3}$), the sphere of influence is bigger than the Hill sphere, making the approximation of the escape conditions less effective as previously explained.  

\item Our analysis is restricted to launch velocities confined to the satellite's orbital plane. 
This assumption simplifies the vector geometry involved in estimating barycentric speeds. 
In the realistic case---such as ejecta generated by an impact---the launch velocities are distributed in three-dimensional space. 
While our energy-based estimates for the residual speed (when ejecta reach the satellite's Hill sphere boundary) remain valid, the direction of the barycentric velocity will depend on any out-of-plane component. 
This vertical component reduces the projection of the residual speed along the satellite's orbital motion. 
For launch directions aligned with the orbital motion, particles benefit less from the satellite’s velocity and are therefore less likely to escape. 
Conversely, if the in-plane component is directed opposite to the satellite’s motion, particles are less hindered and may be more likely to escape. 
As a result, the conditional escape criteria become more restrictive: the upper bound for guaranteed escape is lowered, and the lower bound for possible escape is raised. 
These modified bounds remain  within the range defined by Eq.~(\ref{bounds}), which represent the limiting cases for planar launch. A full three-dimensional treatment to calculate the barycentric velocity, as in \citet{Shute1966}, produces the same bounds as those found here. 

\item
We do not include the full dynamics of the N-body problem, perturbations from other celestial bodies, or effects due to planetary oblateness. 
Additionally, as shown by \citet{Alvarellos2002}, the classical formula for escape speed may overestimate the actual velocity needed to escape from a satellite when the gravitational influence of the host planet is considered. 
For the Moon, this correction reduces the effective escape speed by only about 1\%, but for smaller satellites like Mimas, the correction can be as large as 18\% \citep{Alvarellos2005}. 
Such corrections could affect the thresholds for escape and should be considered in more detailed models.
\end{itemize}

\section{Summary}
\label{sec:Summary}

We developed a patched-conics analytical framework to determine whether ejecta launched from the surface of a satellite can escape the gravitational influence of the planet--satellite system and enter heliocentric orbit, obtaining the following results:
\begin{itemize}
\item We found that two dimensionless factors are key to determine the launch-speed threshold for escape;
the satellite-to-planet mass ratio $\mu$ and the ratio of orbital-to-escape velocity $v_{\text{orb,S}}/ v_{\text{esc,S}}$ (Eq.~\eqref{vl}). 

\item We derived a necessary escape condition based on the Jacobi integral at the $\mathrm{L_2}$ point. This constraint, while less restrictive than the patched-conics escape criterion, yields lower-bound speed estimates (Eq.~\eqref{jacobi_vl}) that are consistent with the sufficient conditions derived earlier. 

\item For most regular satellites of the Solar System planets, the launch speed required for heliocentric escape is 1–2 times (favorable launch conditions) to 5–10 times (unfavorable launch conditions) the satellite's escape speed. 
In contrast, for irregular satellites, the required speed can be hundreds to thousands of times greater than their surface escape speed.

\item Escaping lunar ejecta have unique dynamical behaviors among the Solar System's planetary satellites. 
For the Earth--Moon, its particular satellite-to-planet mass ratio and its low orbital speed relative to lunar surface escape speed leads to a threshold launch speed required for escape to heliocentric orbit that is much lower than for other satellite-planet pairs. In contrast, for most other satellite-planet pairs in the Solar System, barely escaping ejecta will not reach heliocentric orbits.  

\item Since $\mu$ remains constant throughout the Moon’s orbital evolution and the ratio $v_{\text{orb,S}} / v_{\text{esc,S}}$ stays small, the threshold launch speed for heliocentric escape remains low over time. This supports the hypothesis that meteoroids of lunar origin---produced by large impacts in the past---could populate near-Earth and interplanetary space.

\end{itemize}

 \section{Data Availability}

No data was produced in this research. 

\section{Acknowledgments}
We thank two anonymous reviewers for their helpful feedback.
The results reported herein benefited from collaborations and/or information exchange within the program ``Alien Earths'' (supported by the National Aeronautics and Space Administration under Agreement No. 80NSSC21K0593) for NASA’s Nexus for Exoplanet System Science (NExSS) research coordination network sponsored by NASA’s Science Mission Directorate. A.R. acknowledges support by the Air Force Office of Scientific Research (AFOSR) under Grant No. FA9550-24-1-0194.

\printcredits

\bibliographystyle{cas-model2-names}

\bibliography{cas-refs}




\end{document}